\documentclass[twocolumn]{revtex4}
\usepackage{graphicx,natbib,amsmath,amssymb,epsfig,color,psfrag}
\usepackage{fancyhdr}

\input{babarsym}

\newcommand{\dss}{\ensuremath{D^{**}}\xspace}
\newcommand{\ds}{\ensuremath{D^{(*)}}\xspace}

\newcommand{\dspizl}{\ensuremath{D^{(*)}\piz\ell}\xspace}
\newcommand{\Btag}{\ensuremath{B_{\rm tag}}\xspace}

\newcommand{\BDtaunu}{\ensuremath{\Bb \rightarrow D\tau^-\nutb}\xspace}
\newcommand{\BDstaunu}{\ensuremath{\Bb \rightarrow D^*\tau^-\nutb}\xspace}
\newcommand{\BDxtaunu}{\ensuremath{\Bb \rightarrow D^{(*)}\tau^-\nutb}\xspace}
\newcommand{\BDlnu}{\ensuremath{\Bb \rightarrow D\ell^-\nub_{\ell}}\xspace}
\newcommand{\BDslnu}{\ensuremath{\Bb \rightarrow D^{*}\ell^-\nub_{\ell}}\xspace}
\newcommand{\BDxlnu}{\ensuremath{\Bb \rightarrow D^{(*)}\ell^-\nub_{\ell}}\xspace}

\newcommand{\Bpilnu}{\ensuremath{B \rightarrow \pi \ell\nu}\xspace}
\newcommand{\Bzpilnu}{\ensuremath{B^{0} \rightarrow \pi^-\ell^+\nu}\xspace}

\newcommand{\BXlnu}{\ensuremath{B \rightarrow X \ell\nu}\xspace}
\newcommand{\BXclnu}{\ensuremath{B \rightarrow X_c \ell\nu}\xspace}
\newcommand{\BXulnu}{\ensuremath{B \rightarrow X_u \ell\nu}\xspace}

\newcommand{\RDx}{\ensuremath{{\cal R}(D^{(*)})}}
\newcommand{\RDs}{\ensuremath{{\cal R}(D^*)}}
\newcommand{\RD}{\ensuremath{{\cal R}(D)}}

\newcommand{\mmiss}{\ensuremath{m_{\rm miss}^2}\xspace}

\newcommand{\pstarl}{\ensuremath{|\boldsymbol{p}^*_\ell|}\xspace}

\newcommand{\tanB}{\ensuremath{{\rm tan}\beta}\xspace}
\newcommand{\mH}{\ensuremath{m_{H^+}}\xspace}
\newcommand{\tBmH}{\ensuremath{\tanB/\mH}\xspace}

\def\DeltaEDef{\ensuremath{\Delta E = E_{\rm tag} -  \sqrt{s}/2}\xspace}

\begin{document}

\title{Semileptonic $\boldsymbol{B}$ Decays}

\author{Vera G. L\"uth}
\affiliation{SLAC National Laboratory Accelerator, Stanford University, USA}

\begin{abstract}
The following is an overview of the measurements of the CKM matrix elements \Vcb\ and \Vub\
that are based on detailed studies of semileptonic $B$ decays by the  \babar\ and 
Belle Collaborations and major advances in QCD calculations.
In addition, a new and improved measurement of the ratios
$ \RDx = {\cal B}(\BDxtaunu)/{\cal B}(\BDxlnu)$ is
presented. Here $D^{(*)}$ refers to a $D$ or a $D^*$ meson and $\ell$ is either $e$  or $\mu$.
The results, $\RD  = 0.440\pm 0.058\pm 0.042$ and $\RDs = 0.332\pm 0.024\pm 0.018$,
exceed the Standard Model expectations by $2.0\sigma$ and $2.7\sigma$, respectively. 
Taken together, they disagree with these expectations at the $3.4\sigma$ level. 
The excess of events cannot be explained by a charged Higgs boson in the
type II two-Higgs-doublet model.\\
\end{abstract}

\maketitle

\thispagestyle{fancy}

\section{Introduction}

Over the past decade, the vast samples of $B$ mesons recorded at the $B$ Factories at KEK and SLAC have allowed detailed  studies of semileptonic $B$ decays.  In the Standard Model (SM), these decays proceed via first-order weak interaction and are mediated by the W boson.  Decays involving  electrons and muons are expected to be free of non-SM contributions and are therefore well suited for the determination of the Cabibbo-Kobayashi-Maskawa (CKM) matrix elements \Vcb\ and \Vub. They  are fundamental parameters of the SM and have to be determined experimentally. 
Decays involving the higher mass $\tau$ lepton provide additional information on SM processes. They are also sensitive to non-SM contributions, for instance, from the exchange of a charged Higgs boson. 

This presentation will combine a summary of the current status of the measurements of the CKM matrix elements $|V_{cb}|$ and $|V_{ub}|$ from the Belle and \babar\ experiments with the first report on the observation by \babar\ of an excess of events beyond the SM expectations in 
\BDxtaunu\ decays~\cite{convention}.  Here $D^{(*)}$ refers to the ground state charm mesons, $D$ and $D^*$.

\section{$\boldsymbol{\Vcb}$  and $\boldsymbol{\Vub }$} 

There are two experimental methods to determine $|V_{cb}|$ and $|V_{ub}|$. The first is based on the study of exclusive semileptonic $B$ decays in which the hadron is a 
$D$, $D^*$, $D^{**}$,  $\pi$, or $\rho$ meson. The second is based on inclusive decays of the form \BXlnu, where $X$ refers to either $X_c$ or $X_u$, that is, to any allowed hadronic final state with charm or without charm, respectively.
To extract $|V_{cb}|$ or $|V_{ub}|$ from the measured partial decay rates, both approaches depend on calculations of hadronic contributions to the matrix element. 
Since the two methods rely on different experimental techniques and involve different theoretical
approximations, they result in  largely independent measurements of $\Vcb$ and $\Vub$.

The tables summarizing the results from the $B$ Factories are taken from a recent report by the Belle and \babar\ Collaborations~\cite{legacy}. They include updates of input parameter values and reflect the latest understanding of the theoretical uncertainties.
The averages account for correlations among the various measurements. In particular, all theoretical uncertainties are considered to be correlated, as are the uncertainties on the modeling of $\BXclnu$ and $\BXulnu$ decays. Experimental uncertainties due to reconstruction efficiencies are fully correlated for measurements from the same experiment, and uncorrelated for different experiments. Statistical correlations are also taken into account, whenever available. The averaging procedure was developed by the Heavy Flavor Averaging Group (HFAG)~\cite{Amhis:2012bh}.

\subsection{$\boldsymbol{\Vcb}$ from $\boldsymbol{\BDxlnu}$ Decays}

The ''exclusive'' determination of $|V_{cb}|$ relies on studies of \BDlnu\ and \BDslnu\ decays, where $\ell=e,\mu$.
The differential rate for the decay $B\to D\ell\nu$ can be written as
\begin{equation}
    \frac{d\Gamma(B\rightarrow D \ell \nu)}{dw} =
        \frac{G^2_F}{48\pi^3} 
  |V_{cb}|^2  \mathcal{K}_D (w) \, \eta^2_{\rm EW} \,\mathcal{G}^2(w) ,
\label{eq:BtoD}
\end{equation}
\noindent
where $\mathcal{K}_D(w)$ is a known phase-space factor and $\eta_{\rm EW}=1.0066$ refers to the one-loop electroweak correction \cite{Sirlin:1981ie} defined relative to $G_F$ from muon decay.  
In the limit of small lepton mass $m_{\ell}$, 
$\mathcal{G}(w)$ represents a single vector form factor that
depends on the ratio of meson masses $r=m_D/m_B$ and $w=v_B \cdot v_D$, the product 
of the four-velocities of the $D$ and the $B$ mesons.
$w$ is related to the four-momentum transfer $q^2=(P_B-P_D)^2=(P_{\ell}+P_{\nu})^2$, 
namely $w=(m_B^2+m_D^2-q^2)/(2 m_B m_D) = E_D/m_D$.
The values of $w$ are limited by kinematics, $1 \leq w_D \leq 1.59$.

The same ansatz for the differential rate also holds for $B\to D^* \ell \nu$ decays, 
except that the phase-space factor $\mathcal{K}_{D^*}(w)$ and $w$ differ numerically. 
The form factor $\mathcal{G}(w)$ is replaced by $\mathcal{F}(w,\theta_{\ell},\theta_{V},\chi)$, which 
depends also on three angles,  $\theta_{\ell}$ and $\theta_{V}$, the helicity angles of the lepton and the $D^*$, and $\chi$ is the angle between the decay planes of the $D^*$ and the $W$.
$\mathcal{F}(w,\theta_{\ell},\theta_{V},\chi)$ 
contains a combination of three form factors (one vector and two axial vectors) related to the three helicity states of the charm meson. 
The axial vector form factor $A_1(w)$ dominates as $w\to 1$, and therefore  
the decay rate is usually expressed in terms of the ratios $R_1(w)= V(w)/A_1(w)$ 
and $R_2(w)= A_2(w)/A_1(w)$ .

In the limit of infinite b- and c-quark masses, heavy quark symmetry predicts a universal form factor $\mathcal{F}(w)$ with a normalization at zero-recoil,  $\mathcal{F}(w=1)=\mathcal{G}(w=1)$  and a dependence on $w$
which, with constraints from analyticity and unitarity,
can be express in terms of a single parameter $\rho_D^2$ or $\rho^2_{D*}$~\cite{Caprini:1997mu}.

\begin{figure}
 \centering
  \includegraphics[width=6cm]{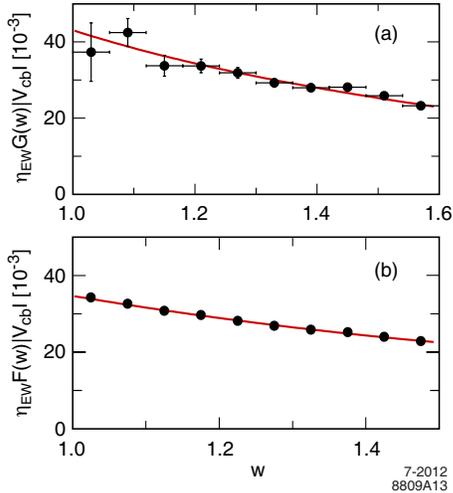}
 \caption{\babar\ measurements,  corrected
    for the reconstruction efficiency, of the $w$ dependence
    of the form factors, with fit results superimposed (solid line):
    (a) $ \eta_{EW} \mathcal{G}(w)\Vcb$ for $B\to D\ell\nu$ 
    decays from tagged events~\cite{Aubert:2009ac}, 
     and for comparison 
    (b) $ \eta_{EW} \mathcal{F}(w)\Vcb$ for $B\to D^*\ell\nu$
    decays from untagged events~\cite{Aubert:2007rs}.
} \label{fig:babar_dlnu}
\end{figure}

The principal uncertainties for the determination of \Vcb\ stem from the form factors, both their shape and  normalization.  
The form factor parameters are determined from fits to the differential decays rates.
For \BDlnu\ decays, $\Vcb\ \mathcal{G}(1)$ and the slope $\rho^2_D$ can be extracted from  $\Gamma(w)$ distribution. For \BDslnu\ decays, $\Vcb\ \mathcal{F}(1)$,  the slope $\rho^2_{D^*}$, $R_1(w=1)$, and $R_2(w=1)$
are determined from fits to the decay distribution $\Gamma(w,\theta_{\ell},\theta_{V},\chi)$.
As an example, Figure~\ref{fig:babar_dlnu} shows the extraction of the form factor slope and normalization from the efficiency-corrected decay rates  for  \BDlnu\ and  \BDslnu\ decays for two \babar\ analyses.

The results of recent  form factor measurements for \BDlnu\ and \BDslnu\ decays from Belle and \babar\ are listed in Table~\ref{tab:Dlnu}.   
The branching fractions quoted for $B^0$
are based on $B^0$ and $B^+$ measurements, combined under the assumptions that isospin relations hold.
The branching ratios are calculated using these form factor
parameters, taking into account correlated systematic uncertainties.
The errors are dominated by the uncertainties in the detector performance.

For \BDslnu\ decays, only two of the four measurements exploit the angular dependence of the form factor  $\mathcal{F}(w,\theta_{\ell},\theta_{V},\chi)$.
The most precise measurement based on the full Belle data set
~\cite{Dungel:2010uk} relies on a  fit to the 1-dimensional distributions of the four variables. 
\babar~\cite{Aubert:2007rs} enhances the sensitivity to $R_1(1)$ and $R_2(1)$ and also $\Vcb$ by combining the results with a fit to the four-dimensional decay rate
$\Gamma(w,\theta_{\ell},\theta_V, \chi)$~\cite{Aubert:2006cx}.
The results from the two experiments agree well. The average values 
$R_1(1)=1.40\pm 0.03 \pm 0.01$  and  $R_2(1)=0.86\pm 0.02 \pm 0.01$, have a precision of  3\%, and are
used as input to measurements that are limited to the $w$ dependence of the decay rate. 
 
\begin{table*}
  \centering
  \caption{Summary of the $B$~Factory results for the \BDlnu\ and \BDslnu\ 
    form factor parameters $\eta_\mathrm{EW}\mathcal{G}(1)\Vcb$,
    $\eta_\mathrm{EW}\mathcal{F}(1)\Vcb$, $\rho^2_{D}$ and 
    $\rho^2_{D^*}$, and the branching fractions. The measurements have been
    rescaled to the end of year 2011 values of the common input
    parameters \cite{Beringer:1900zz,Amhis:2012bh}. The stated errors correspond to the statistical 
    and systematic uncertainties.} 
  \begin{tabular}{lccc}
    \hline \hline
    \BDlnu\ & $\eta_\mathrm{EW}\mathcal{G}(1)\Vcb$ (10$^{-3}$) &
    $\rho^2_D$ & $\mathcal{B}(B^0 \to D^-\ell^+\nu)$  (\%) \\
    \hline
    Belle~\cite{Abe:2001yf} & $40.8\pm 4.4\pm 5.0$ 
    & $1.12\pm 0.22\pm 0.14$ & $2.07\pm 0.12\pm 0.52$ \\
    \babar\ $DXl\nu$~\cite{Aubert:2008yv} & $43.4\pm 0.8\pm 2.1$
    & $1.20\pm 0.04\pm 0.06$ & $2.18\pm 0.03\pm 0.13$\\
    \babar\ tagged~\cite{Aubert:2009ac} & $42.5\pm 1.9\pm 1.1$ 
    & $1.18\pm 0.09\pm 0.05$ & $2.12\pm 0.10\pm 0.06$ \\
    \hline
    Average & $42.7\pm 0.7\pm 1.5$ 
    & $1.19\pm 0.04\pm 0.04$ & $2.14\pm 0.03\pm 0.10$\\
    \hline \hline

    \\
    \hline \hline
    \BDslnu & $\eta_\mathrm{EW}\mathcal{F}(1)\Vcb$ (10$^{-3}$) &
    $\rho^2_{D^*}$ & $\mathcal{B}(B^0\to D^{*-}\ell^+\nu)$ (\%) \\
    \hline
     \babar\ $D^{*-}\ell^+\nu$~\cite{Aubert:2007rs} & $34.1\pm 0.3\pm 1.0$ 
    & $1.18\pm 0.05\pm 0.03$ & $4.58\pm 0.04\pm 0.25$ \\
    \babar\ $\bar D^{*0}e^+\nu$~\cite{Aubert:2007qs}& $35.1\pm 0.6\pm 1.3$ 
    & $1.12\pm 0.06\pm 0.06$ & $4.95\pm 0.07\pm 0.34$ \\
    \babar\ $DXl\nu$~\cite{Aubert:2008yv} & $35.8\pm 0.2\pm 1.1$ 
    & $1.19\pm 0.02\pm 0.06$ & $4.96\pm 0.02\pm 0.20$\\
   Belle~\cite{Dungel:2010uk} & $34.7\pm 0.2\pm 1.0$ 
    & $1.21\pm 0.03\pm 0.01$ & $4.59\pm 0.03\pm 0.26$ \\
    \hline
    Average & $35.5\pm 0.1\pm 0.5$ 
    & $1.20\pm 0.02\pm 0.02$ & $4.83\pm 0.01\pm 0.12$\\
    \hline \hline
   \label{tab:Dlnu}
  \end{tabular}
\end{table*}

A precise determination of \Vcb\ requires corrections to $\mathcal{G}(1)$ and $\mathcal{F}(1)$ for finite quark masses and non-perturbative effects. 
Table~\ref{tab:vcbexcl} summarizes the latest results from lattice QCD (LQCD), heavy quark sum rules (HQSR), and HQE calculation.  The LQCD predictions for the two decay modes are about 5\% higher than the results from the other two QCD calculations.

\begin{table}
  \centering
  \caption{Normalization of the form factors for \BDlnu\ and \BDslnu\ decays
   and the resulting values of \Vcb\ based on different QCD calculations.
   } 
  \label{tab:vcbexcl}
  \begin{tabular}{lcc}
    \hline \hline
    \BDlnu\ & & \\
    Calculation & $\eta_\mathrm{EW}\mathcal{G}(1)$ & $ \Vcb\ (10^{-3}$) \\ \hline
    LQCD~\cite{Okamoto:2004xg}
    & $ 1.081 \pm 0.018 \pm 0.016 $ & $ 39.46 \pm 1.54 \pm 0.88$\\
    HQE~\cite{Uraltsev:2003ye}

    & $  1.047 \pm 0.020 $ & $ 40.79 \pm 1.58 \pm 0.78$ \\
        \hline \hline
     \\
      \hline \hline
   \BDslnu\ & & \\
    Calculation & $\eta_\mathrm{EW}\mathcal{F}(1)$ & $ \Vcb\ (10^{-3})$\\ \hline
    LQCD~\cite{Bernard:2008dn,Bailey:2010gb}
    & $ 0.908 \pm 0.005 \pm 0.016 $ & $ 39.04 \pm 0.55 \pm 0.73$\\
    HQSR~\cite{Gambino:2010bp}
    & $ 0.865 \pm 0.020 $ & $40.93 \pm 0.58 \pm 0.95$ \\
    \hline \hline
  \end{tabular}
\end{table}

While the results for the two decay modes agree well, $\Vcb$ measured
in $B\to D^*\ell\nu$ decays is more precise and will be considered as
the main result. 
The differences in the values for \Vcb\ underline the fact that the  
principal uncertainties  stem from the form factor normalization. 

\subsection{$\boldsymbol{\Vcb}$  from Inclusive $\boldsymbol{\BXclnu}$ Decays}

At the parton level, the inclusive decay rate for $b\to c \ell \nu$
can be calculated accurately: It is proportional to $|V_{cb}|^2$ and
depends on the quark masses $m_b$ and $m_c$.  To extract $|V_{cb}|$
from the measured $B$ meson decay rate, the parton-level calculations
have to be corrected for effects of strong interactions. 
These corrections are suppressed by powers of $\alpha_s$ and
$\Lambda_{QCD}/m_b$ ($\Lambda_{QCD}$ refers to the perturbative QCD
scale). 
Thus, the decay rate for inclusive semileptonic $B$ decays can be 
expressed in terms of a Heavy Quark Expansion (HQE) 
in inverse powers of the $b$-quark mass and 
with a limited number of non-perturbative parameters. 
Due to confinement and non-perturbative effects,
HQEs rely on the definition of the quark masses, which depends on the coupling to the SM Lagrangian.  

Calculations of the decay rates for $B \to X_c \ell \nu$ are available in the $1S$ mass scheme~\cite{Hoang:1998hm,Bauer:2004ve}
which is derived from a perturbative expression for the mass of the $Y(1S)$ state, and the kinetic mass scheme~\cite{Bigi:1994ga,Benson:2003kp} which is derived from heavy quark sum rules and enters the non-relativistic expression for the kinetic energy of the $b$ quark inside the $B$ meson. 
In the following, we rely on calculations in the kinetic scheme
for which the total decay rate for $B \to X_c \ell \nu$ can be expressed to ${\cal O}(1/m_b^3)$ in a simplified way as 
\begin{eqnarray}
\label{eq:opesl}
\nonumber
\Gamma_{c \ell \nu} &\cong& \frac{G_F^2 m_b^5}{192\pi^3} |V_{cb}|^2 (1+A_{ew}) A_{pert}(r,\mu)  \\
&\times&\Bigg [ z_0(r) 
               + z_2(r) \Bigg (\frac{\mu_{\pi}^2}{m_b^2},\frac{\mu_G^2}{m_b^2}\Bigg )
               + z_3(r) \Bigg (\frac{\rho_D^3}{m_b^3},\frac{\rho_{LS}^3}{m_b^3} \Bigg )\Bigg] .\nonumber \\
\end{eqnarray}
\noindent
A more detailed ansatz can be found in the literature~\cite{Benson:2003kp}.
The dependence on the charm quark mass $m_c$ is contained in the ratio 
$r=m_c/m_b$ which enters the phase space factors $z_i(r)$.  
The most relevant scale for  $b\rightarrow c$ transitions is the energy release $m_b-m_c$. 
The electroweak corrections are estimated to be $1+A_{ew} \approx [ 1 + \alpha/\pi \, \mathrm{ln}(M_Z/m_b) ]^2 \approx 1.014 $ 
and the perturbative contributions are $A_{pert}\approx 0.91 \pm 0.01$. 
The  leading non-perturbative contributions arise at ${\cal O}(1/m_b^2)$ and are parameterized in terms of $\mu_{\pi}^2(\mu)$ and $\mu_{G}^2(\mu)$, the expectation values of the kinetic and chromomagnetic dimension-five operators. At ${\cal O}(1/m_b^3)$, two additional parameters enter, $\rho_{D}^3(\mu)$ and $\rho_{LS}^3(\mu)$, the  expectation values 
of the Darwin and spin-orbital (LS) dimension-six operators. These parameters, as well as the quark masses $m_b$ and $m_c$, depend on the renormalization scale $\mu$ which separates short-distance from long-distance QCD effects. For the kinetic scheme the chosen value is $\mu=1~\gev$~\cite{Bigi:1996si}.
 
Similar HQEs can be derived for moments of inclusive $B \to X_c \ell \nu$ distributions; they also depend on $\alpha_s$, $m_b$ and $m_c$  and the same perturbative parameters.  
The leading terms
are known  to ${\cal O} (\alpha_s)$ and ${\cal O} (\alpha_s^2 \beta_0)$ (with
$\beta_0=(33-2n_f)/3$). Non-perturbative terms are included to ${\cal O}(1/m_b^3)$, and corrections to ${\cal O} (\alpha_s^2)$ have recently been implemented.

Moment measurements are available from $B \to X_c \ell \nu$ decays for the lepton energy spectrum $\langle E_{\ell}^n \rangle$ with $n=1,2,3$, the hadronic mass distribution, $\langle M^{2n}_X \rangle$ with $n=1,2,3$ 
These moments and the inclusive semileptonic decay rate $\Delta {\cal B}$  are measured for different values of the minimum 
lepton energy $E^{min}_{\ell}$ in the range of $0.6 - 1.5 \gev$.

The HFAG has developed a fit procedure based on
the full $\mathcal{O}(\alpha^2_s)$ calculations of the moments in the
kinetic scheme~\cite{Gambino:2011cq,Gambino:2011fz}. 
This fit combines moment measurements from the $B$ Factories and determines $\Vcb$, 
the $b$-quark mass $m_b$, and four higher order
parameters in the OPE description of semileptonic decays. The only
external input is the average $B^0$ and $B^+$~lifetime, $(1.582\pm 0.007)$ ps~\cite{Beringer:1900zz}.

\begin{figure} 
 \centering
  \includegraphics[width=7.5cm]{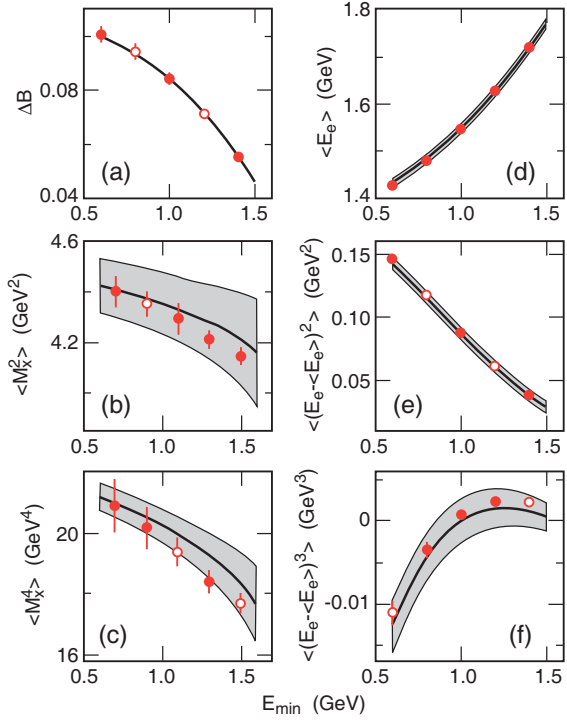}
\caption{Comparison of the measured moments and the fit to the HQE predictions (kinetic scheme) by the Belle
Collaboration~\cite{Schwanda:2008kw} as a function of the minimum lepton or photon energy:
(a) branching fraction $\Delta B$, (b,c)  hadron mass ($M_X$), (d,e,f) and electron energy 
($E_e$). The vertical bars represent the experimental uncertainties and the shaded bands show the theoretical uncertainties.  Filled (open) circles mark data used (unused) in the fit. 
}
\label{fig:Vcbglobal}
\end{figure}

Figure~\ref{fig:Vcbglobal} shows the comparison of fitted HQE predictions in the kinetic scheme to some of the moments measured by the Belle Collaboration~\cite{Schwanda:2008kw} as a function of the minimum lepton energy.  
Since the moments are derived from the same distribution, in particular those which differ only by the minimum  lepton energy, are highly correlated, only about half of the measured data points are included in the fit. Similar fits have been performed by the BABAR collaboration, including also some mixed moments of different distributions, e.g.,  a combination of the hadronic mass and lepton energy~\cite{Aubert:2009qda}. 

The fit to selected 44 moments measured by Belle and \babar\ has a $\chi^2/NDF=23.2/37$, an indication that the experimental and theoretical uncertainties and estimated correlations among moments are not fully understood.  
Correlations between the fitted parameters are generally small.

To enhance the precision on $m_b$,  a precise constraint on the $c$-quark mass $m_c(3~\mathrm{GeV})=0.998\pm 0.029$~\gev, was introduced, derived from low-energy sum rules~\cite{Dehnadi:2011gc}, one of several precise calculations of quark masses~\cite{Chetyrkin:2009fv,McNeile:2010ji}. 
Fits in the kinetic scheme to the Belle and \babar\ moments  result in
\begin{eqnarray}
 |V_{cb}|_{\rm incl} &=& (42.01\pm 0.47_{fit} \pm 0.59_{th}) \times 10^{-3},  \nonumber \\
 m^{kin}_b(1\gev)   &=& (4.551\pm 0.025_{fit}) \gev,                    \nonumber \\
 \mu^{2 kin}_{\pi}(1\gev) &=& (0.499\pm 0.044_{fit}) \gev^2.                     
\end{eqnarray}
\label{eq:opefit}
The first error represents the combined experimental and theoretical uncertainty of the fit, and the additional 
error on \Vcb\  reflects the estimated uncertainty of 1.4\% for the expansion for the decay rate.  The fitted semileptonic branching fraction is ${\cal B} (B\to X_c\ell\nu)=(10.51 \pm 0.13)\%$.  The result on \Vcb\ cited here  agrees very well with the result of a fit to the same moments based on the 1S mass scheme~\cite{legacy}.

\subsection{$\boldsymbol{\Vub}$  from $\boldsymbol{\Bpilnu}$ Decays}

For the determination of $|V_{ub}|$ from exclusive charmless decays, the most promising decays, both experimentally and theoretically, is $\Bbar \to \pi \ell^- \nub_{\ell}$.  Branching fractions for decays involving 
the pseudoscalar mesons $\eta$ and $\eta'$ and the vector mesons $\rho$ and $\omega$ have been measured, albeit with considerable uncertainties. Thus, they currently provide only limited information on form factors and therefore on \Vub.

As for \BDlnu\ decays,
the differential decay rate for decays to low-mass charged leptons can be written as 
\begin{equation}
    \frac{d\Gamma(B^0\rightarrow \pi^- \ell^+ \nu)}{dq^2} =
        \frac{G^2_F}{24\pi^3} |p_{\pi}|^3 
        |V_{ub}|^2  |f_+(q^2)|^2 .
\label{eq:Btopi}
\end{equation}
\noindent
Here $f_+(q^2)$ is the only form factor affecting the rate,
because in the limit of small lepton masses $m_{\ell}$, the term proportional to the second form factor $f_0(q^2)$ can be \
neglected~\cite{Gilman:1989uy}. 
Since the rate depends on the third power of $p_{\pi}$, the pion momentum in the $B$ meson rest frame, it is suppressed at high $q^2$. 

Among several parameterizations of the form factors, a model-independent approach based on the general properties of analyticity, unitarity and crossing-symmetry is preferred~\cite{Boyd:1994tt,Becher:2005bg}. 
The stringent constraints on the form factor
are expressed in the form of a rapidly converging series in the variable 
$z(q^2,q^2_0) = (\sqrt{m^2_{+} - q^2} - \sqrt{m^2_{+} - q^2_0} )$, with $m_{\pm}=m_B \pm m_{\pi}$. 
The simplest functional form by Bourreley, Caprini, and Lellouch (BCL)~\cite{Bourrely:2008za} is
\begin{equation}
\label{eq:BCL}
  f_+(q^2) = \frac{1}{1-q^2/m^2_{B^*}}\sum_{k=0}^{K} b_k(q^2_0) z(q^2,q^2_0)^k .
\end{equation}
Here $q^2_0$ is a free parameter, chosen to optimize the convergence. 

The principal goal of the studies of $\Bbar \to \pi \ell^- \nub_{\ell}$ decays is a precise
measurement of the branching fraction and the determination of the $q^2$
dependence of the $B\to\pi$ form factor.
The main experimental challenge is the reduction of the 
 abundant background from continuum events and from \BXclnu\ decays. Also the isolation 
of the \Bpilnu decays
from the other \BXulnu decays, where $X_u$ is a charmless hadronic final state,
is difficult due to very similar decay kinematics.

Three analyses have been performed based on untagged event samples, two by  the
\babar~\cite{delAmoSanchez:2010uj,delAmoSanchez:2010zd} and one by the Belle~\cite{Ha:2010rf}
Collaboration.  The measured branching fractions show excellent agreements, the average, taking into account correlations, 
${\cal B}(\Bzpilnu) = (1.44 \pm 0.03 \pm 0.05)\time 10^{-4}$, is dominated by systematic uncertainties, primarily related to the reconstruction of the missing neutrino derived from the missing energy and momentum in the event, and the backgrounds from continuum events at low $q^2$ and from $B \to X_u \ell \nu$ decays at high $q^2$.

\begin{figure}
\centering
\includegraphics[width=6.0cm]{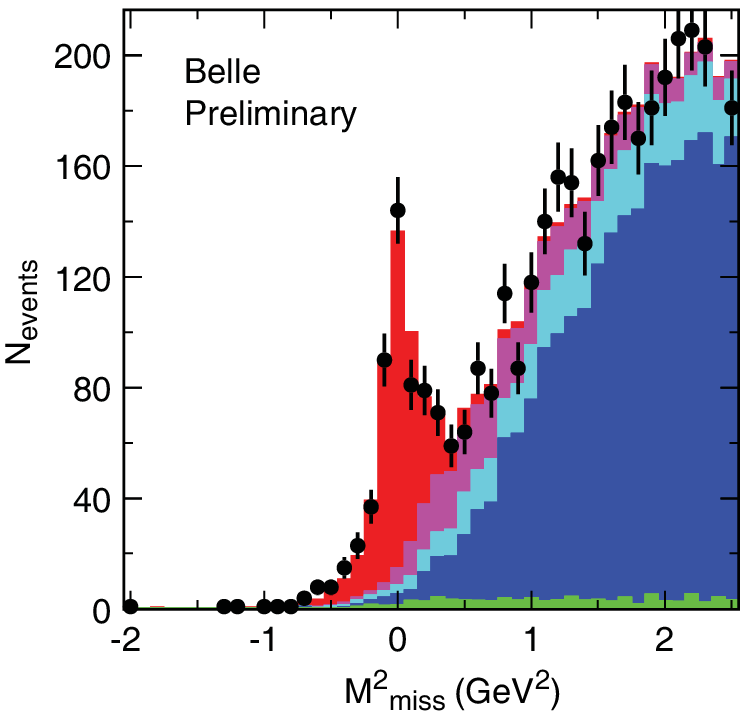}  
\caption{Missing mass squared distributution from a tagged samples of \Bzpilnu\ decays
by the Belle Collaboration~\cite{belle_pilnu}.}
\label{fig:Vub_Excl_belle}
\end{figure}

A few months ago the Belle collaboration~\cite{belle_pilnu} presented first results on  exclusive charmless decays involving the pseudoscalar mesons $\pi^+, \pi^0, \eta,$ and $\eta '$ and the vector mesons, $\rho^+, \rho^0,$ and $ \omega$. This analysis is based on the full data sample and benefits from a highly efficient selection of events tagged by the hadronic decay of one of the $B$ mesons in the event. Figure~\ref{fig:Vub_Excl_belle} shows the missing mass distribution for a
selected \Bzpilnu\ sample with a purity of about 65\%.  From a fit to this distribution, a signal of $468 \pm 28$ \Bzpilnu\  decays  has  been extracted.  Preliminary measurements of the $d\Gamma/dq^2$ distributions and branching factions
are fully consistent with the untagged measurements (see Table~\ref{tab:vub}).  While the statistical uncertainties are larger than for the untagged analyses, the systematic uncertainties are much reduced due to the  kinematic reconstruction of the full event. 

Currently,  two principal methods are used to extract $|V_{ub}|$ from the measured differential decay rates. The more conventional method relates the measured partial branching fractions to
$\Delta \zeta(q^2_{min},q^2_{max})=\Delta \Gamma_{theory} / |V_{ub}|^2$, which is derived from QCD calculations integrated over specific $q^2$ ranges, and
\begin{equation} 
\Vub^2 = \Delta{\cal B}(q^2_{min},q^2_{max})/\Delta\zeta(q^2_{min},q^2_{max}) / \tau_0  ,
\end{equation}
where $\tau_0$ is the $B^0$~lifetime. Table~\ref{tab:vub} lists the average partial branching fractions, the values of $\Delta\zeta$, and the $|V_{ub}|$ results relying on light cone sum rules (LCSR), and two sets of LQCD calculations.

More recently, $|V_{ub}|$ has been determined from a simultaneous fit to unquenched LQCD calculations~\cite{Bailey:2008wp} and the measured $q^2$ spectrum.  The BCL parameterization
is used as parameterization for $f_+(q^2)$ over the whole $q^2$ range
to minimize the form factor model dependence. 
This method makes optimum use of the measured shape of the whole $q^2$ 
spectrum and normalization  from LQCD which results in a reduced uncertainty on \Vub .

\begin{table*}[tbp]
 \renewcommand{\arraystretch}{1.4}
  \centering
  \caption{Overview of \Vub measurements based on \Bpilnu decays (for 3 untagged and 1 tagged samples) for 
    various $q^2$ regions and form factor calculations: 
    LCSR~\cite{Khodjamirian:2011ub}, HPQCD~\cite{Dalgic:2006dt},
    FNAL/MILC~\cite{Bailey:2008wp}. The quoted errors on \Vub are due to
    experimental uncertainties and theoretical uncertainties on $\Delta\zeta$. 
    The last column shows the \Vub results of the simultaneous fits
    to data and the FNAL/MILC prediction. Here the stated error
    represents the combined experimental and theoretical
    uncertainty.} \label{tab:vub}
  \begin{tabular}{lcccc}
    \hline\hline
    &  LCSR                           & HPQCD
    & FNAL/MILC                       & FNAL/MILC fit          \\
    \hline
    $\Delta\zeta$ (ps$^{-1}$) & $4.59^{+1.00}_{-0.85}$          &
    $2.02{\pm}0.55$                 & $2.21^{+0.47}_{-0.42}$
    & $2.21^{+0.47}_{-0.42}$ \\
    $q^2$ range ($\gev^2$)      & $0-12$                          &
    $16-26.4$                       & $16-26.4$
    & $16-26.4$              \\
    \hline
    Experiment                  & \multicolumn{4}{c}{$\Vub$
      ($10^{-3}$)} \\
    \hline
    \babar\ (6 bins)~\cite{delAmoSanchez:2010uj}
    & $3.54 \pm 0.12^{+0.38}_{-0.33}$ 
    & $3.22 \pm 0.15^{+0.55}_{-0.37}$ 
    & $3.08 \pm 0.14^{+0.34}_{-0.28}$
    & $2.98 \pm 0.31$        \\
    \babar\ (12 bins)~\cite{delAmoSanchez:2010zd}
    & $3.46 \pm 0.10^{+0.37}_{-0.32}$ 
    & $3.26 \pm 0.19^{+0.56}_{-0.37}$ 
    & $3.12 \pm 0.18^{+0.35}_{-0.29}$
    & $3.22 \pm 0.31$        \\
    Belle~\cite{Ha:2010rf}                      
    & $3.44 \pm 0.10^{+0.37}_{-0.32}$ 
    & $3.60 \pm 0.13^{+0.61}_{-0.41}$ 
    & $3.44 \pm 0.13^{+0.38}_{-0.32}$
    & $3.52 \pm 0.34$        \\
    \hline
    \babar/Belle untagged              
    & $3.47 \pm 0.06^{+0.37}_{-0.32}$ 
    & $3.43 \pm 0.09^{+0.59}_{-0.39}$ 
    & $3.27 \pm 0.09^{+0.36}_{-0.30}$
    & $3.23 \pm 0.30$        \\
    \hline
    Belle tagged~\cite{belle_pilnu}               
    & $3.38 \pm 0.17^{+0.37}_{-0.31}$ 
    & $3.86 \pm 0.25^{+0.53}_{-0.53}$ 
    & $3.69 \pm 0.24^{+0.39}_{-0.35}$
    &  ---      \\
    \hline\hline
  \end{tabular}

\end{table*}

\begin{figure}
\centering
\includegraphics[width=7cm]{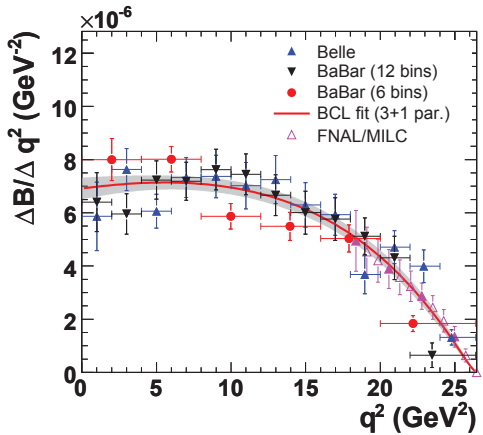}  
\caption{Simultaneous fit of the BCL parameterization
to the $\Delta{\cal B}/\Delta q^2$ distributions for 
$B\to\pi\ell\nu$ decays
and to four of the twelve points of the
FNAL/MILC calculation (magenta, closed triangles)
The FNAL/MILC prediction has been rescaled to the data according to
the \Vub\ value obtained in the fit.}
\label{fig:Vub_Excl}
\end{figure}

Figure~\ref{fig:Vub_Excl} shows the combined fit to the 
FNAL/MILC lattice calculations and the data from the
three untagged measurements.  
To avoid high correlations, only four of the twelve
FNAL/MILC points have been included in the fit.
This reduction of the theoretical input does not change the \Vub result but
leads to a better agreement of the fitted curve with the lattice points.
The $\chi^2$ probability of the fit is $2.2\%$ ($\chi^2/ndf = 58.9/31$). 
The fit results for the parameters in the BCL parameterization are
$b_1/b_0 = -0.82 \pm 0.20$ and 
$b_2/b_0 = -1.63 \pm 0.62$, 
and a value of $f_+(0)\Vub = 0.945 \pm 0.028$ is obtained, 
which translates to $f_+(0) = 0.29 \pm 0.03$,
in good agreement with the LCSR result, $f_+(0) = 0.28 \pm 0.02$.
The \Vub values obtained from fits to the individual untagged measurements 
agree with each other within about one standard deviation. 
The total uncertainty on \Vub\ is about $9\%$; 
$3\%$ from the branching fraction measurement, $4\%$ from the
shape of the $q^2$ spectrum determined with data, and $8\%$ from the
form-factor normalization obtained from LQCD.

Table~\ref{tab:vub} summarizes various measurements of \Vub, based on 
different form factor normalizations. In addition to the three untagged analysis, 
it also lists the preliminary results by Belle using the new tagging algorithm.
All these results are fully consistent within the stated uncertainties.

\subsection{$\boldsymbol{\Vub}$  from Inclusive $\boldsymbol{\BXulnu}$ Decays}

The total inclusive rate for \BXulnu\  decays can be expressed in terms of an OPE 
which has a similar structure as the one for \BXclnu\ decays,
with nonperturbative corrections occuring  at $O(1/m_b^2)$ and higher.
Unfortunately, experimenters usually restrict the phase space to reduce 
the large background from Cabibbo-favored \BXclnu\ decays, and these restrictions 
spoil the HQE convergence. Perturbative and non-perturbative corrections 
are drastically enhanced and the rate becomes sensitive to the Fermi motion
 of the $b$ quark inside the $B$ meson, introducing terms that are not 
suppressed by powers of $1/m_b$. In practice, non-perturbative 
shape functions (SF) are introduced, which to leading order in $1/m_b$ should 
be similar for $b\to u$ and $b\to s$ transitions.  The form of 
the SF cannot be calculated from first principles, but has to be constrained 
by data.  SF parameterizations are generally chosen such that their first 
and second moments are equal to $\overline{\Lambda}=m_B - m_b$ and $\mu^2_{\pi}$, 
i.e., they are directly related to the non-perturbative HQE parameters and 
thus can be determined by fits to moments of $B\to X_c \ell \nu$ and $B\to X_s \gamma$ 
decays.

\begin{figure}
 \centering
\includegraphics[width=8cm]{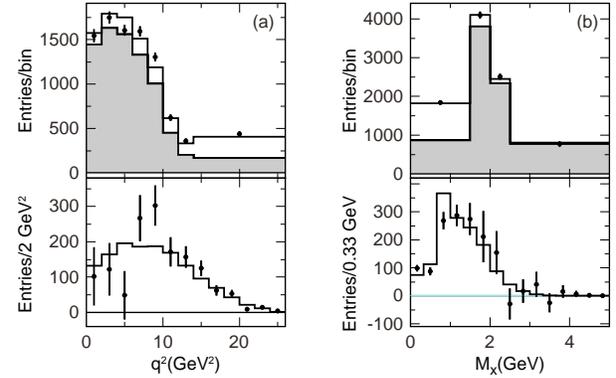} 
\caption{\babar~\cite{Lees:2012vv}: 
Projections of measured distributions (data points) of (a) $q^2$ and (b) $M_X$ 
for inclusive $B\to X_u\ell\nu$ decays, Upper row:  
comparison with the result of a $\chi^2$ fit to the two-dimensional 
$M_X-q^2$ distribution for the sum of two scaled MC contributions, signal (white) 
and background (grey).  Lower row: corresponding spectra
  with equal bin size after background subtraction based on the
  fit. The data are not efficiency corrected.} 
\label{fig:Vub_Incl}
\end{figure}

\begin{table*}  
  \centering
  \caption{Overview of \Vub measurements based on inclusive $B\to
    X_u\ell\nu$ decays analyzed in three untagged and 2 tagged data samples. The
    critical input parameters $m_b$ and $\mu^2_{\pi}$ depend on the
    different mass schemes and have been obtained from the OPE fits to $B\to
    X_c\ell\nu$ hadronic mass moments in the kinetic
    mass scheme. For the BLNP and DGE calculations, they have been
    translated from the kinetic to the shape function and
    $\overline{MS}$ schemes, respectively. The additional
    uncertainties on $m_b$ and $\mu^2_{\pi}$  are due to these scheme
    translations. The first error is the experimental and the second
    reflects the uncertainties of the QCD calculations and the HQE
    parameters~\cite{Amhis:2012bh}.} 
  \label{tab:vubincl}
  \begin{tabular}{lccc}
    \hline\hline
    &   BLNP              & GGOU              & DGE  \\ 
    \hline            
    Mass scheme            &  SF scheme          &  Kinetic scheme   
    & $\overline{\rm MS}$ scheme\\
    $m_b$ (GeV) & $4.588 \pm 0.023 \pm 0.011$ & $4.560 \pm 0.023$ 
    & $4.194 \pm 0.043 $\\
    $\mu^2_{\pi}$ (GeV$^2$) & $0.189^{+0.041}_{-0.040} \pm 0.020$ &
    $0.453 \pm 0.036$ & --- \\
    \hline
    Experiment & & $\Vub$ (10$^{-3}$) & \\
    \hline
    Belle \cite{Limosani:2005pi} 
    & $ 4.88\pm 0.45 ^{+0.24}_{-0.27} $ & $ 4.75 \pm 0.44 ^{+0.17}_{-0.22} $ 
    & $ 4.79\pm 0.44^{+0.21}_{-0.24}  $ \\
    \babar \cite{Aubert:2005mg} 
    & $ 4.48\pm 0.25 ^{+0.27}_{-0.28} $ & $ 4.29 \pm 0.24 ^{+0.18}_{-0.24}$ 
    & $ 4.28\pm 0.24 ^{+0.22}_{-0.24} $ \\ 
    \babar \cite{Aubert:2005im} 
    & $ 4.66\pm 0.31 ^{+0.31}_{-0.36} $ & ---                           
    & $ 4.32 \pm 0.29^{+0.24}_{-0.29}        $ \\
    \hline
    Average untagged                
    & $ 4.65\pm 0.22 ^{+0.26}_{-0.29}  $ & $ 4.39\pm 0.22 ^{+0.18}_{-0.24} $ 
    & $ 4.44\pm 0.21 ^{+0.21}_{-0.25} $\\
    \hline
    Belle \cite{Urquijo:2009tp} 
    & $ 4.47\pm 0.27 ^{+0.19}_{-0.21} $ & $ 4.54 \pm 0.27^{+0.10}_{-0.11} $ 
    & $ 4.60\pm 0.27 ^{+0.11}_{-0.13} $\\
    \babar \cite{Lees:2012vv} 
    & $ 4.28\pm 0.24 ^{+0.18}_{-0.20}$ & $ 4.35 \pm 0.24 ^{+0.09}_{-0.11} $ 
    & $ 4.40 \pm 0.24^{+0.12}_{-0.13} $\\
    \hline
    Average tagged 
    & $ 4.35\pm 0.19 ^{+0.19}_{-0.20}$ & $ 4.43\pm 0.21 ^{+0.09}_{-0.11} $ 
    & $ 4.49 \pm 0.21 ^{+0.13}_{-0.13}$\\
    \hline \hline
  \end{tabular}
\end{table*}

The extracted values of \Vub\ are presented in Table \ref{tab:vubincl}
for both untagged and tagged \BB\ samples.
The earlier untagged measurements placed cuts near the kinematic limit 
of the lepton spectrum.  Thus they 
covered limited fractions of the total phase space and had sizable 
experimental and theoretical uncertainties.   
The most recent analyses by the Belle~\cite{Urquijo:2009tp} and
\babar~\cite{Lees:2012vv} Collaborations are based on their full data sample and
use \BB\ events tagged by the hadronic decays of the second $B$ meson and thus 
cover up to 90\% of the phase space. 
The most precise results are based on a fit to the two-dimensional $q^2$ 
versus $M_X$ (the mass of the $X_u$ system) distributions. 
An example of such a fit is shown in Figure \ref{fig:Vub_Incl}. 
The average of the two partial branching fraction measurements, assuming full
correlation of the uncertainty in the predicted signal spectrum, is
$\Delta {\cal B} (p^*_{\ell} >1\gev) = (1.87 \pm 0.10 \pm 0.11) \times
10^{-3}$.
Here $p^*_{\ell}$ refers to the momentum of the charged lepton in the rest frame of the $B$ meson.  
The systematic uncertainties are
dominated by the simulation of the signal decays; in particular, they
are sensitive to the shape function and the $b$-quark mass.

There is a high degree of consistency among the measurements and the results 
for different QCD calculations show little variation. 
 Based on results in Table~\ref{tab:vubincl}, we quote the unweighted arithmetic
 average of the tagged data samples as the overall result,
\begin{equation}
  \Vub_\mathrm{incl} = (4.42\pm 0.20_\mathrm{exp}\pm 0.15_\mathrm{th})\times 10^{-3}.
\end{equation}

\subsection{Summary on $\boldsymbol{\Vcb}$ and $\boldsymbol{\Vub}$ }

As a result of joint efforts by theorists and experimentalists, our understanding of semileptonic $B$-meson decays has substantially advanced over the last decade.  

Substantial progress has been made in the application of HQE
calculations to extract \Vcb\ and $m_b$ from fits to measured moments
from $\BXclnu$ decays. The total error quoted on $\Vcb$ is 1.8\% and the introduction of a c-quark mass constraint, $m_c(3\gev) = (0.998 \pm
0.029)\gev$, has reduced the overall uncertainty on $m_b$ to 25~\mev.

The measurement of $\Vcb$ based on the exclusive decay $B\to D^*\ell \nu_{\ell}$
has now a combined experimental and theoretical uncertainty
of 2.3\%. 
Values of \Vcb\ differ by about 5\%,
depending on the choice of the QCD calculation for the normalization
of the form factors; lattice calculations lead to lower values of
\Vcb\ than heavy flavor sum rules.

Consequently the comparison of the inclusive and exclusive
determinations of \Vcb\ depends on the choice of the normalization of
the form factors. For the LQCD calculations, the values of the
inclusive and exclusive determination of \Vcb\ differ at the level of
$2.5\sigma$, 
\begin{eqnarray}
 \Vcb_\mathrm{excl} & = & [39.04 \ (1 \pm 0.014_\mathrm{exp} \pm
   0.019_\mathrm{th})] \times 10^{-3}~,   \nonumber \\
 \Vcb_\mathrm{incl} & = & [42.01 \ (1 \pm 0.011_\mathrm{exp} \pm
   0.014_\mathrm{th})] \times 10^{-3}~. \nonumber
\end{eqnarray}

On the other hand, based on heavy flavor sum rule calculations for the exclusive measurement, the value is  
\begin{equation}
 \Vcb_\mathrm{excl} = [40.93 \ (1 \pm 0.014_\mathrm{exp} \pm
   0.023_\mathrm{th})] \times 10^{-3} \,  \nonumber \\
\end{equation}
\noindent
and agrees very well with the inclusive measurement. 

For inclusive  measurements of \Vub , experimental and theoretical errors are comparable in size. The dominant experimental uncertainties are related to the limited size of the tagged samples, the signal simulation, and background subtraction. 
The theoretical uncertainties are dominated by the error on the 
$b$-quark mass; a 20-30\mev uncertainty in $m_b$ impacts \Vub\ by 2-3\%.  

Measurements of the differential decay rate as a function of $q^2$ for 
\Bpilnu\ provide valuable information on the shape of
the form factor, though with sizable errors due to large
backgrounds. Results based on different QCD calculations agree within
the stated theoretical uncertainties. While the traditional method of
normalizing to QCD calculations in different ranges of $q^2$ results
in uncertainties of $^{+17\%}_{-10\%}$, combined fits to LQCD
predictions and the measured spectrum using a theoretically motivated
ansatz \cite{Boyd:1994tt,Becher:2005bg,Bourrely:2008za} have resulted
in a reduction of the theoretical uncertainties to about 8\%.

The values of the inclusive and exclusive determinations of \Vub\ are
only marginally consistent, they differ at a level of $3\sigma$,
\begin{eqnarray}
  \Vub_\mathrm{excl} & = & [3.23 \ (1 \pm 0.05_\mathrm{exp} \pm 0.08_\mathrm{th})] \times 10^{-3} \,   \nonumber \\
  \Vub_\mathrm{incl} & = & [4.42 \ (1 \pm 0.045_\mathrm{exp}\pm 0.034_\mathrm{th})] \times 10^{-3}.  \nonumber 
\end{eqnarray}

\section{Study of $\boldsymbol{\BDxtaunu}$ Decays}

So far, we have focused on semileptonic decays involving low-mass charged leptons,
for instance,  \BDxlnu\ decays which are well-understood SM processes. Decays involving the higher mass $\tau$ lepton  provide an opportunity to search for contributions beyond the SM processes, 
for example, decays mediated by a charged Higgs boson of the Two Higgs Doublet Model (2HDM) of type II~\cite {Grzadkowski:1992qj,Tanaka:1994ay,Itoh:2004ye,Nierste:2008qe,Tanaka:2010se,Fajfer:2012vx}.

In the SM, the differential decay rate (integrated over angles) for \BDxtaunu\ decays can be written in terms of helicity amplitudes as follows
~\cite{Korner:1989qb,Falk:1994gw,Hwang:2000xe}, 

\begin{eqnarray}
 \frac{d\Gamma_\tau}{dq^2}
  &=& \frac{G_F^2|V_{cb}|^2|{\rm\bf p}|q^2}{96\pi^3m_B^2}
      \biggl[1 - \frac{m_\tau^2}{q^2}\biggr]^2
      \biggl(\left[|H_{\scriptsize +}|^2 + |H_{\scriptsize  -}|^2\right. \nonumber \\
  && \left. + |H_{0}|^2\right]
      \biggl[1 + \frac{m_\tau^2}{2q^2}\biggr]
   + \frac32\frac{m_\tau^2}{q^2}|H_{s}|^2 \biggr),
\end{eqnarray}
where for simplicity, the $q^2$ dependence of the helicity amplitudes $H_{n}$ has been omitted.  The amplitudes $H_{\pm}$ only receive contributions from helicity $\lambda_{D^*}=\pm$ and therefore are absent for \BDtaunu\ decays.  $\lambda_{D^*}=0$ contribute to $H_{0}$ and 
$H_{s}$.  %
SM calculations~\cite{Kamenik:2008tj,Fajfer:2012vx}, updated to account for recent FF measurements, predict for the ratios of decay rates,
\begin{eqnarray}
 {\cal R}(D)_{\rm SM}   &=&  \frac{\Gamma(\bar B \to D\tau\nu)}{\Gamma(\bar B \to D\ell\nu)}
=0.297 \pm 0.017, \nonumber \\
 {\cal R}(D^*)_{\rm SM} &=& \frac{\Gamma(\bar B \to D^*\tau\nu)}{\Gamma(\bar B \to D^*\ell\nu)} =0.252 \pm 0.003. \nonumber
\end{eqnarray}
These ratios are independent of \Vcb\ and to a large extent, insensitive to the parameterization of the hadronic matrix element.  Previous measurements 
~\cite{Matyja:2007kt,Aubert:2007dsa,Bozek:2010xy}
have slightly exceeded these predictions, though due to sizable statistical uncertainties the significance of the measured excess was low. 

The \babar\ Collaboration reports results~\cite{Lees:2012xj} of a major update of its earlier measurement~\cite{Aubert:2007dsa} of the ratios \RDx\ for both charged and neutral $B$ mesons.
They choose to reconstruct only the leptonic decays $\tau^- \to \ell^- \nub_{\ell} \nu_{\tau}$, so that the final states of the signal \BDxtaunu\ and the normalization \BDxlnu\ decays contain the same particles, {\it i.e.,} a charm meson $D^{(*)}$ and  a charged lepton, $e^-$ or $\mu^-$.  This leads to a cancelation of various experimental uncertainties in the ratios \RDx. 

The analysis relies on the reconstruction of the full \BB\ final state.  In addition to the semileptonic decay, the hadronic decay of the other $B$ meson is fully reconstructed.Compared to the previous \babar\ analysis~\cite{Aubert:2007dsa}, 
the efficiency of the $B$ tagging algorithm and 
the  reconstruction of the semileptonic decays has been increased by a factor of three, 
and the size of the data sample is doubled.

The events are divided into four subsamples identified by the 
charm meson from a semileptonic decay candidate,  $D^0\ell, D^{*0}\ell, D^+\ell, D^{*+}\ell$. The missing mass
$\mmiss=(p_{\epem} - p_{\rm tag} - p_{D^{(*)}} - p_{\ell})^2$ separates the normalization decays with $\mmiss\sim 0$  (one neutrino) from signal decays with much larger \mmiss (three neutrinos). The leptons in normalization decays have higher momenta
than the secondary leptons from $\tau$ decays, and this feature is also utilized to separate the two decay modes.
Decays to higher-mass, excited charm mesons, $B\to {D^{**}\ell/\tau\nu}$ , enter the 
event selection, whenever the low momentum pion from  
$\dss\to\ds\pi$ decays is undetected or incorrectly assigned to the \Btag. 
These higher-mass states are poorly understood and their branching fractions are not well measured. Therefore  the fit includes four control $D^{(*)} \piz \ell$ samples, enriched in 
$B\to {D^{**}\ell/\tau\nu}$ decays, by  adding a \piz\ decay to the signal selection. 

\begin{figure}
\psfrag{x}[Br]{\footnotesize{\pstarl (GeV)}} 
\psfrag{D0}[cl]{\footnotesize{$D$}}
\psfrag{Ds0}[cl]{\footnotesize{$\Dstar$}}
\includegraphics[width=8cm]{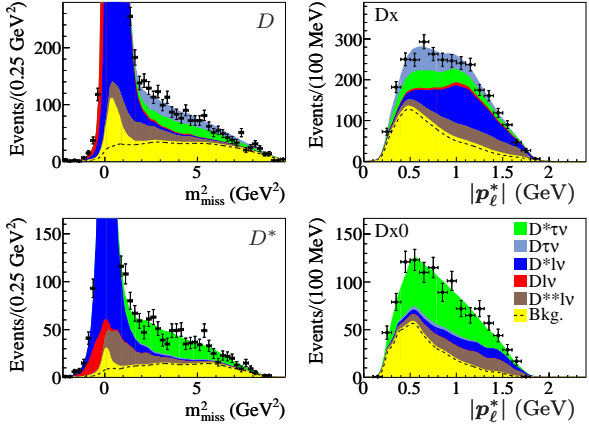}
\caption {\mmiss and \pstarl projections of the isospin-constrained fit to the signal samples. 
The \pstarl projections do not include the \mmiss peak at $\mmiss<1\gev^2$, 
which excludes most of the normalization events~\cite{Lees:2012xj}.
} 
\label{fig:Fit_Signal} 
\end{figure}

The background in these 8 samples is reduced by applying multivariate methods (BDT) that make use of  variables describing the quality of the reconstruction,
such as the mass of the reconstructed \ds and \DeltaEDef, where $E_{tag}$ and $\sqrt{s}$ refer to the $B_{tag}$ and the center of mass energy,
respectively.  Candidates with one or more additional charged tracks are eliminated.  
The yields for semileptonic decay of the four signal \BDxtaunu\ decays and four
normalization \BDxlnu\ decays are extracted by an extended, unbinned maximum-likelihood fit 
to the two-dimensional \mmiss\ versus \pstarl\ distributions.
The fit is performed simultaneously to the four $\ds\ell$ samples and the four \dspizl samples. 
The distribution of each $\ds\ell$ sample is described as the sum of eight contributions: 
$D\tau\nu$, $\Dstar\tau\nu$, $D\ell\nu$, $\Dstar\ell\nu$, $\dss\ell\nu$,  cross-feed between \Bz\ and \Bp due to misreconstruction of the $B_{tag}$, and backgrounds from \BB\ 
and continuum events. The yields and shapes of these backgrounds are taken from MC simulation and fixed in the fit. 
A large fraction of $B\to D^{*}\ell\nu$ decays are reconstructed in the 
$D\ell\nu$ samples. A total of 56 two-dimensional probability density functions (PDF) for the individual contributions 
to the samples are constructed from large Monte Carlo samples by using Gaussian Kernel Estimators (KEYS).  
The fitted distributions for $D \ell$ and $D^*\ell$ samples are shown in Figure~\ref{fig:Fit_Signal}.  The results for charged and neutral $B$ mesons are combined, assuming isospin relations.  In total, there are $489 \pm 63$ \BDtaunu\ compared to $2,981 \pm 65$ \BDlnu\ decays,  and $888\pm 63$ \BDstaunu\ compared to $11,953 \pm 122$ \BDslnu\ decays. 

The measured ratios, corrected for efficiencies and branching fractions,
are
\begin{eqnarray}
 {\cal R}(D)   &=& 0.440 \pm 0.058 \pm 0.042, \nonumber\\
 {\cal R}(D^*) &=& 0.332 \pm 0.024 \pm 0.017. \nonumber 
\end{eqnarray}
The principal contributions to the  systematic errors are the uncertainties in the $B\to {D^{**}\ell/\tau\nu}$  and continuum backgrounds and the limited size of the MC samples. 
The measurements exceed the SM calculations by $2.0 \sigma$ and $2.7\sigma$. The combination of these results, taking into account their correlation
of -0.27,  excludes the SM at the $3.4 \sigma$ level.

The charged Higgs boson $H^+$ would only impact the helicity amplitude $H_{s}$,
\begin{equation}
 H_{s}^{\rm 2HDM} = H_{s}^{\rm SM}
                     \left[1 - \frac{m_b\, q^2 }{m_b \mp m_c}
                               \frac{\tan^2\beta}{m_{H}^2} \right].
\end{equation}
\noindent
The negative sign applies to \BDtaunu\ and the positive sign to \BDstaunu\ decays.
Depending on the value of $\tan \beta/\mH$,  the ratio of two vacuum expectations values and the mass of the charged Higgs, this term would either enhance or decrease the ratios \RDx\ and affect the $\tau$ polarization.
 
\begin{figure}
\psfrag{R\(D\)}[bl]{\RD}
\psfrag{R\(D*\)}[bl]{\RDs}
\psfrag{t}[Br]{\tBmH (GeV$^{-1}$)}
\includegraphics[width=7.0cm]{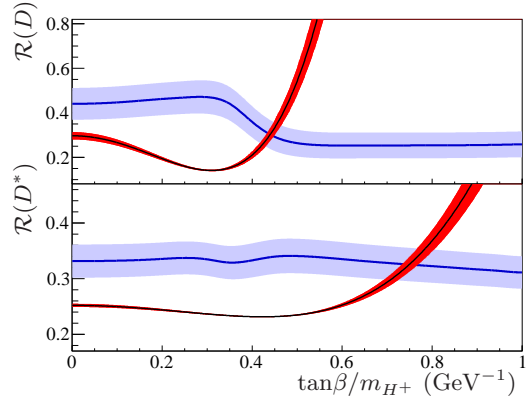}
\caption {Comparison of the results of the \babar\ analysis (light grey, blue)~\cite{Lees:2012xj}
with predictions that include a charged Higgs boson of type II 2HDM (dark grey, red).  The SM corresponds to $\tanB/\mH=0$. }
\label{fig:PRL_Higgs}
\end{figure}

Figure \ref{fig:PRL_Higgs} shows the impact a charged 2HDM type II Higgs boson~\cite{Tanaka:2010se,Barger:1989fj} would have on the measured ratios \RD\ and \RDs\ as a function $\tan \beta/\mH$.  
This assessment was made by reweighting the simulated events to account for the changes in the matrix element, including the $\tau$ polarization.

The measured values of \RD\ and \RDs\ match the predictions of this particular Higgs model
for  $\tanB/\mH=(0.44\pm0.02)\gev^{-1}$ and $\tanB/\mH=(0.75\pm0.04) \gev^{-1}$, respectively.
The  \RD\ and \RDs\ results together exclude the type II 2HDM charged Higgs boson at at 
99.8\% confidence level, or higher for larger values of $\tanB/\mH$. 
This conclusion is only valid for values of \mH\ greater than $15\gev$~\cite{Tanaka:1994ay,Tanaka:2010se}.  However, 
the region for $\mH\leq 15\gev$ has already been excluded by $B\to X_s\gamma$ measurements \cite{Misiak:2006zs}, and therefore, the type II 2HDM is excluded in the full  \tanB--\mH parameter space.

\section{Conclusions and Outlook}

While there has been tremendous progress, we have not achieved the
precision of 1\% for \Vcb\ or 5\% on \Vub, goals many of us had hoped to reach before the shutdown of Belle and \babar\ experiments.

We are left with two puzzles: 
\begin{itemize}
\item
The puzzling difference, in the results of exclusive and inclusive measurements of \Vub\ and to lesser extent of \Vcb, if we rely on non-lattice calculations, which challenge our current understanding of the experimental and theoretical techniques applied.
\item
The excess of events in \BDxtaunu\ decays at the level of 3.4 standard deviations relative to the SM calculations, which  might indicate non-SM contributions.  This excess cannot be explained by contributions from a 
charged Higgs boson of the 2HDM of type II.  However,  it has been pointed out in recent publications~\cite{Fajfer:2012jt,Crivellin:2012ye}
that this result can be accommodated in terms of other versions of the Two-Higgs Doublet Model.
\end{itemize}

To resolve these puzzles a major effort will be required. 
It will take much larger tagged data samples and a more detailed
understanding of the detector performance and background composition to
reduce experimental uncertainties. It will also require further progress in
QCD calculations, based on lattice, heavy flavor sum rules, or other
methods, to reduce the uncertainties of form factor predictions for
exclusive decays,  to improve the detailed prediction of inclusive
processes, and to incorporate precision determinations of the  heavy
quark masses.

Measurements of the $D^*$ and $\tau$ polarization and forward-backward asymmetries as well as other kinematic distributions might be able to distinguish among various couplings of non-SM processes~\cite{Datta:2012qk,Becirevic:2012jf} and  possibly lead to an explanation of the excess events in \BDxtaunu\ decays. 
\bigskip

\begin{acknowledgments}
The author would like to thank the organizers of FPCP 2012, in particular 
Zheng-guo Zhao and his colleagues at USTC in Hefei for a very exciting conference.  The summary of the results on \Vcb\ and \Vub\ is heavily based on the {\it Physics of the $B$ Factories}, a book that has been assembled by the \babar\ and Belle Collaborations with the support of many theorists. I consider myself  privileged to have been part of this effort.  Last not least, I would like to acknowledge the contibutions of Manuel Franco Sevilla, who recently completed his PhD thesis on the \BDxtaunu\ decays and the observation of an excess of events above Standard Model expectations.

This work was supported by Department of Energy contract DE-AC03-76SF00515.
\end{acknowledgments}

\bigskip 

\begin{thebibliography}{66}
\bibitem[{con()}]{convention}
\bibinfo{note}{Throughout this report, charge-conjugate decay modes are
  implied}.
\end{thebibliography}


\end{document}